\documentclass[%
 aip,
 jmp,%
 amsmath,amssymb,
preprint,%
]{revtex4-1}

\usepackage{subfigure}
\usepackage{graphicx}
\graphicspath{{./}{./images/}{./images_all/}}
\DeclareGraphicsExtensions{.eps}

\usepackage{dcolumn}
\usepackage{bm}

\begin{document}


\title
{An error-resilient non-volatile magneto-elastic universal logic gate with ultralow energy-delay product}

\author{Ayan K. Biswas$^{1}$, Jayasimha Atulasimha$^2$
and Supriyo Bandyopadhyay}
\email{sbandy@vcu.edu}
\affiliation{
$^1$Department of Electrical and Computer Engineering, Virginia Commonwealth University,
Richmond, VA 23284, USA \\
$^2$Department of Mechanical and Nuclear Engineering, Virginia Commonwealth University,
Richmond, VA 23284, USA}
\date{\today}

\maketitle

{\bf
A long-standing goal of computer technology is to process
and store digital information with the {\it same} device in order to
implement new architectures.
One way to accomplish this is to use
nanomagnetic `non-volatile' logic gates that can perform Boolean operations and then
store the output data in the magnetization states
of nanomagnets, thereby doubling as both logic and
memory.
Unfortunately, many proposed nanomagnetic gates
do not possess the seven essential
characteristics
of a Boolean logic gate : concatenability, non-linearity,
isolation between input and output,
gain, universal logic implementation, scalability and error resilience.
More importantly, their energy-delay products and error-rates vastly exceed that of
conventional transistor-based logic gates, which is a drawback. Here, we
propose a non-volatile voltage-controlled nanomagnetic logic gate
 that possesses all the necessary characteristics of a logic gate and
 whose energy-delay product is $\sim$2 orders of magnitude {\it less} than that
of other nanomagnetic (non-volatile) logic gates and $\sim$1 order of magnitude 
less than that of (volatile) CMOS-based logic gates. The error-resilience is
also superior to that of other known nanomagnetic gates.}

There is significant interest in `non-volatile logic' because the
ability to store and process information with the same device  affords
immense flexibility in designing computing architectures. Non-volatile
logic based architectures  can reduce overall energy dissipation by eliminating
refresh clock cycles, improve system
reliability and
produce `instant-on' computers with virtually no boot delay.
 A number of
non-volatile universal logic gates have been proposed to date
\cite{cowburn, ney, behin-aein}, but they do
not necessarily satisfy all the requirements for a logic gate \cite{behin-aein2, Waser}
and therefore may not be usable in all circumstances. Ref.
 [\onlinecite{cowburn}]
proposed an idea where digital bits are stored in the magnetization orientations
of an array of dipole-coupled nanomagnets and dipole coupling between neighbors
elicits logic operation on the bits. This gate is not concatenable since the
input and output
bits are encoded in dissimilar physical quantities: the inputs are encoded
in directions of
magnetic fields and the output is encoded in the magnetization orientation of
a nanomagnet. Thus, the output of a preceding gate cannot act as the input to the
succeeding gate without additional transducer hardware to convert
the magnetization orientation of a nanomagnet into the direction of a magnetic field.
The gate also lacks true gain since the energy needed to switch the output comes from
the inputs and not an independent source such as a power supply. Additionally, the
strength of dipole coupling between magnets decreases as the square of the magnet's
volume, which limits scalability. Finally, dipole coupling is not sufficiently
resilient
against thermal noise, resulting in unacceptably large dynamic bit error
rate in dipole-coupled logic gates \cite{roychowdhury,salehi,bokor}.

Ref. [\onlinecite{ney}] proposed a different construct where a NAND gate was implemented
with a single magneto-tunneling junction
(MTJ) placed close to four current lines, two of which are input lines, the third is required for an initialization operation,
and the fourth is the output line. The input bits are coded in the {\it directions} of the 
currents in the two input lines, while the output bit is coded in the {\it magnitude} of the 
current in the output line. The input currents generate a magnetic field (whose 
direction is determined by the directions of the input currents) which then orients the magnetization 
of the MTJ's soft layer in the direction
of the field and
 determines the MTJ resistance as well as the magnitude of  the output current.
The magnitude of the output current was shown to be a NAND function of the directions of 
the input currents \cite{ney}. Slightly different renditions of this idea
have been proposed \cite{lee} and an experimental demonstration of MTJ-based logic has been reported \cite{jpwang}.
Unfortunately, this gate too is not directly concatenable since the input bits are encoded in the
{\it directions} of the input currents while the output bit is encoded in the
{\it magnitude}
of the output current. Moreover,
since it is difficult to confine magnetic fields to small regions,
the separation between neighboring devices must be large. Individual devices
can be small in size, but because the inter-device pitch is large,
the device density will be small.
There is also some chance that the output current can, by itself, switch the magnetization
of the soft magnetic layer and therefore affect its own state. This is equivalent to
lack of isolation between the input and the output, which makes gate operation unreliable.
Finally, another MTJ-based logic gate has been proposed recently \cite{kuntal_new}, but it requires a feedback circuit to operate
(which makes it energy-inefficient and error-prone) and even the 
logic functionality is questionable \cite{comment}.
Thus, while these devices are interesting
in their own right, they may
not be universally usable.

A more recent scheme that overcomes most of the above shortcomings was proposed in
Refs. [\onlinecite{behin-aein}] and [\onlinecite{behin-aein2}]. It implements non-volatile logic with
magnets
switched by spin currents. Both computation and communication between
gates are carried out with a sequence of clock pulses. Unfortunately, its error-resilience
 has not been examined. Normally, magnetic devices
are much more error-prone than transistors since magnetization dynamics is
easily disrupted by thermal noise \cite{roychowdhury,salehi}. Logic
has stringent requirements on error rates and  it is imperative to
evaluate the dynamic bit error probability of any gate to assess its viability.

Finally, the most important metric for a logic gate is the energy-delay
 cost. All non-volatile magnetic logic schemes are deficient in this area.
The scheme in Ref. [\onlinecite{ney}]
uses current-generated magnetic fields to switch magnets and hence
would dissipate at least 10$^9$ kT of energy per gate operation at room temperature
to switch in $\sim$1 ns \cite{salehi2}  (energy-delay product =
4$\times$10$^{-21}$ J-s).
A recent experiment conducted to demonstrate switching of nanomagnets
with  on-chip current-generated magnetic fields ended up dissipating
approximately
10$^{12}$ kT of energy per switching event, despite switching in $\sim$1 $\mu$s
(energy-delay product = 4$\times$10$^{-15}$ J-s) \cite{nd}.
  The scheme in
Ref. [\onlinecite{behin-aein}] is expected to dissipate between 10$^5$ and
10$^6$ kT of
energy
when it switches in 1 ns
(energy-delay product = 4$\times$10$^{-25}$ - 4$\times$10$^{-24}$ J-s) \cite{sharad}, although a lower
energy-delay product may be possible with design optimization \cite{sarkar}.
 On the other hand, a low-power CMOS transistor is claimed to dissipate only
 10$^3$ kT of energy  when it switches
in 0.1 ns (energy-delay product = 3$\times$10$^{-28}$ J-s)
\cite{nikonov}, although more realistic estimates based on available data
predict energy dissipation of $\sim$450 aJ to switch in 0.34 ns
(energy-delay product = 1.5$\times$10$^{-25}$ J-sec) \cite{Datta-review}.
Since any non-mainstream technology must eclipse mainstream CMOS technology by at least close to an
order of magnitude to be worthy of consideration, the magnetic logic ideas have languished despite their
coveted non-volatility.

In this report, we propose a non-volatile nanomagnetic NAND gate that is switched with voltage
(not current) unlike the other schemes. It has
 an energy-delay product 2.78$\times$10$^{-26}$ J-s
 which is smaller than that of other magnetic logic schemes
by almost two orders of magnitude and that of CMOS by almost one order of magnitude.
The energy-delay product however, by itself, is not the most meaningful metric for benchmarking device performance.
It is always possible to reduce this product arbitrarily by sacrificing reliability. For example, one can forcibly switch a
device faster and also dissipate less energy to switch (which will reduce the energy-delay product), but at the cost
of increased switching failures. A more meaningful metric may be the product of energy, delay and failure (error) probability.
The error
probability for the proposed NAND gate
has been evaluated rigorously from stochastic simulations.
With careful choice of parameters, it is
possible to reduce the error probability to below 10$^{-8}$
at room temperature, which is remarkable for
magnetic logic (magnetic logic is typically much more error-prone than transistor logic). Finally, the proposed gate fulfills {\it all} the requirements for logic.
Therefore, it is possibly the first nanomagnetic logic gate that has
the cherished advantage of magnetic logic gates (non-volatility) and yet none of the usual
disadvantages.

The proposed gate structure is shown in Fig. 1(a). It is
implemented with a skewed MTJ stack, passive resistors $R$ and $R_P$,  a bias dc voltage $V_{BIAS}$,
and a constant current source $I_{BIAS}$. The current source is not used to switch the gate, but merely to produce
an output voltage $V_{out}$ representing the output logic bit. Input bits are encoded in
input voltages $V_{in}$. Both input and output bits are encoded in the same physical
quantity, voltage, which allows direct concatenation.

The bottom layer of the MTJ stack is an elliptical magnetostrictive (metallic) nanomagnet (Terfenol-D) and the top layer
is a non-magnetostrictive elliptical (metallic) synthetic anti-ferromagnet (SAF) with large shape anisotropy.
The top layer acts as the hard (or pinned) layer and the bottom layer acts as the soft (or free) layer of
the MTJ.
There is a small permanent magnetic field directed along the minor axis of the
magnetostrictive nanomagnet (+y-direction)
which brings its two stable magnetization orientations out of the ellipse's major axis
 and aligns them along two mutually
perpendicular in-plane directions that lie between the
major and minor axes (Fig. 1(b)) \cite{tiercelin,Giordano}. The major axis
of the top SAF layer is aligned collinear with one of the two stable magnetization orientations
of the soft magnet. It is then permanently magnetized in the direction {\it anti-parallel} to that stable
orientation.  As a result, when the magnetization of the soft layer is in this stable orientation,
the hard and soft layers have anti-parallel magnetization resulting in high MTJ resistance.
When the soft layer's magnetization is in the other stable direction, the MTJ resistance is lowered.

\begin{figure}
\includegraphics[width=5.1in]{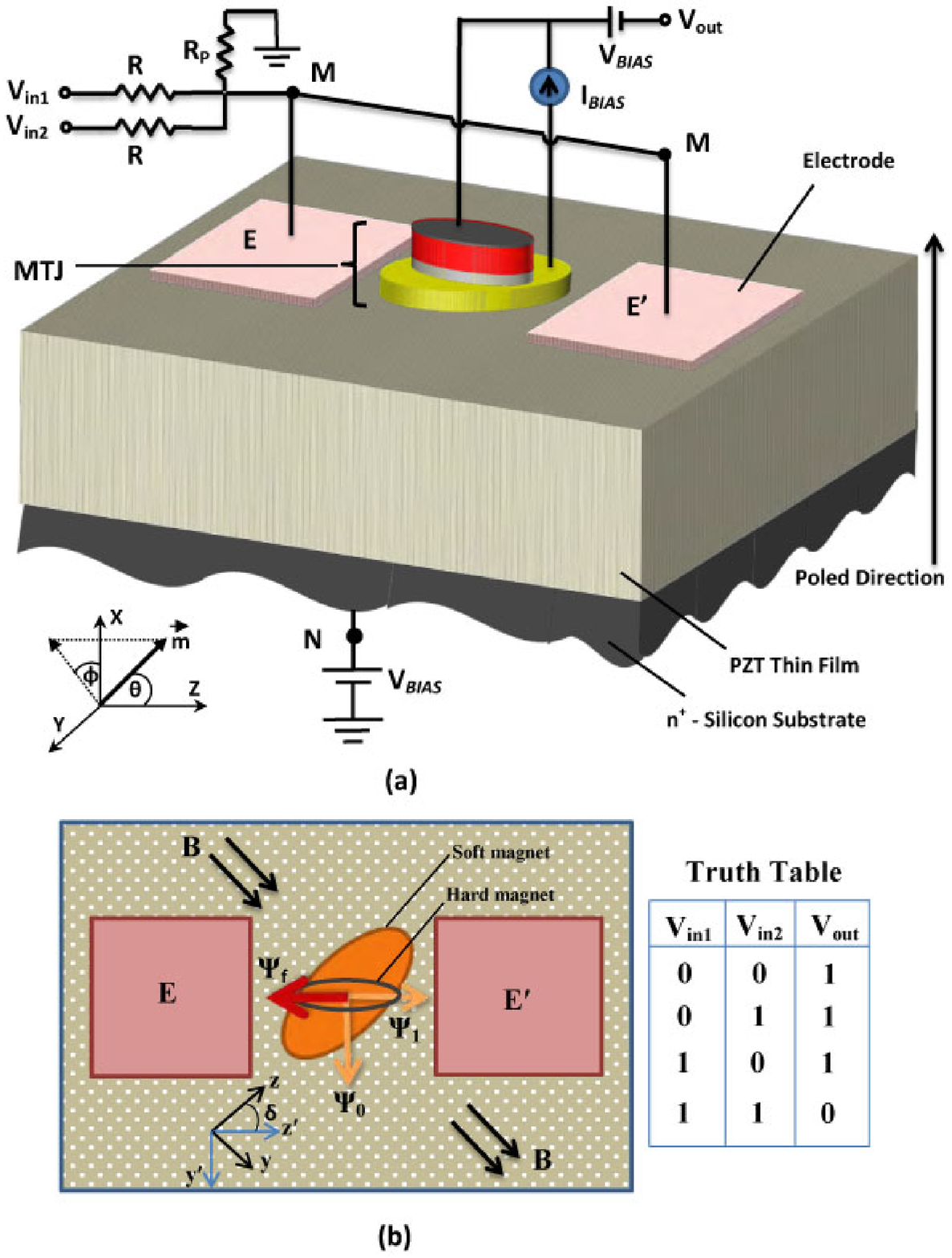}
 \caption[Caption of figure]{{\bf Structure of a NAND gate}. (a) The PZT film has a thickness of
$\sim$50 nm and is deposited on a conducting n$^+$-Si substrate. It is poled with an electric field
in the direction shown. The distance between the electrodes is 80 nm
and the electrode lateral dimensions are also of the same order. (b)
The fixed magnetization
orientation of the top (hard) magnet is denoted by $\Psi_f$, and the two stable magnetization orientations
of the bottom (soft) magnet are denoted by $\Psi_0$ and $\Psi_1$. The MTJ resistance is high when the soft magnet's
magnetization is aligned along $\Psi_1$. The MTJ resistance is (ideally)
a factor of
2 lower when the soft magnet's
magnetization is aligned along $\Psi_0$. The slanted ellipse is the footprint
of the soft magnet
and the horizontal ellipse is the footprint of the hard magnet. The black double arrows show the direction of
the permanent magnetic field.
}
  \label{fig:Fig1}
\end{figure}

Two electrodes
$E$ and $E'$ are delineated on the PZT surface such that the line joining their centers is collinear with
the major axis of the hard layer and hence also the first stable orientation of the soft layer.
The electrode lateral dimensions, the separation between their edges, and the PZT film thickness are all
approximately equal.
These two electrodes are electrically shorted. Whenever an electrostatic potential difference appears
between them and the underlying silicon substrate (between point-$M$ and point-$N$ in Fig. 1(a)), the PZT layer is strained.
Since the electrode in-plane dimensions are comparable to the PZT film thickness, the out-of-plane ($d_{33}$)
expansion/contraction and the in-plane ($d_{31}$) contraction/expansion of the piezoelectric regions underneath
the electrodes
produce a highly localized strain field under the electrodes \cite{Lynch2013}. Furthermore, since the
electrodes are separated by a distance approximately equal to the
PZT film thickness, the interaction between the local strain fields below the electrodes
will lead to a biaxial strain in the PZT layer underneath the soft magnet \cite{Lynch2013}.
This biaxial strain (compression/tension
along the line joining the electrodes and tension/compression along the perpendicular axis) is transferred to
the soft magnetostrictive magnet in elastic contact with the PZT, thus rotating its magnetization via the Villari effect.
This happens despite any substrate clamping and despite the
fact that the electric field in the PZT layer just below the magnet is approximately zero \cite{Lynch2013}. Some of the generated strain may even reach the top hard magnet \cite{li}, but since
the hard magnet is very anisotropic in shape and is not magnetostrictive, its magnetization will not rotate
perceptibly. Rotation of the magnetization of the soft layer of an MTJ due to strain has been
recently demonstrated experimentally \cite{li}.

Fig. 2 shows the potential energy profile of the soft magnetostrictive nanomagnet in its own plane ($\phi$ = 90$^{\circ}$)
plotted as a function of the
angle $\theta$ subtended by its magnetization vector with the major axis of the ellipse (z-axis).
Note
that the energy profile has two degenerate minima ($B$ and $C$)
in the absence of stress (i.e. when no voltage is applied
between nodes $M$ and $N$). These two states correspond to directions $\Psi_1$ and $\Psi_0$, respectively, in Fig. 1,
which are the two stable magnetization orientations of the soft layer with $\Psi_1$ being anti-parallel to
the magnetization of the hard layer and $\Psi_0$ being approximately perpendicular to $\Psi_1$.
Application of sufficient potential difference between $M$ and $N$, to generate
sufficient stress in the magnetostrictive magnet,
transforms the energy profile into a
monostable well (with no local minima)
located at either $B$ or $D$, depending on whether
the stress component along the line EE' is tensile or compressive, i.e. whether node $M$ is at a higher
potential than node $N$, or the opposite \cite{tiercelin,Giordano}. If we apply compressive
stress with the right voltage polarity, the system will go to point $D$ and the magnetization
will point along the corresponding direction very close to $\Psi_0$. Thereafter, if we
withdraw the voltage and stress, the system will go to the {\it nearer} energy minimum at point $C$
(and not the other minimum at $B$) because
of the
potential barrier that exists between  $B$ and $C$. This happens with $>$99.999999\% probability at
room temperature in the presence of thermal noise (see supplementary material).
Once it reaches $C$, the system will remain there (since it is an energy minimum) and the magnetization
will continue to point along the corresponding direction $\Psi_0$
(making the device non-volatile) until tensile stress is applied [by applying voltage of opposite polarity between
$M$ and $N$] to take the system to
$B$, thereby changing the magnetization to the
other stable direction $\Psi_1$. Upon withdrawal of the tensile stress, the system will remain in state $B$
because the energy barrier between $B$ and $C$ will prevent it from migrating to $C$. Therefore, the system is
non-volatile in either state.
By merely choosing the {\it polarity} of the voltage between nodes $M$ and $N$, we can
deterministically visit either state $B$ or state $C$ and orient the magnetization along
either of the two stable states $\Psi_1$ and $\Psi_0$. In other words, by applying voltages of two
different polarities, we can make the MTJ resistance either high or low.
The soft nanomagnet's magnetization (and hence the MTJ resistance)  will remain in the chosen state after
the voltage is withdrawn. This was used as the basis for
deterministically writing the bit 0 or 1 in non-volatile memory, irrespective of what the initial
stored bit was \cite{tiercelin,Giordano}. Here, we have extended that idea to build a non-volatile
universal logic gate (NAND) using a magneto-tunneling junction in the manner of Ref. [\onlinecite{ney}].

\begin{figure}
\includegraphics[width=6.5in]{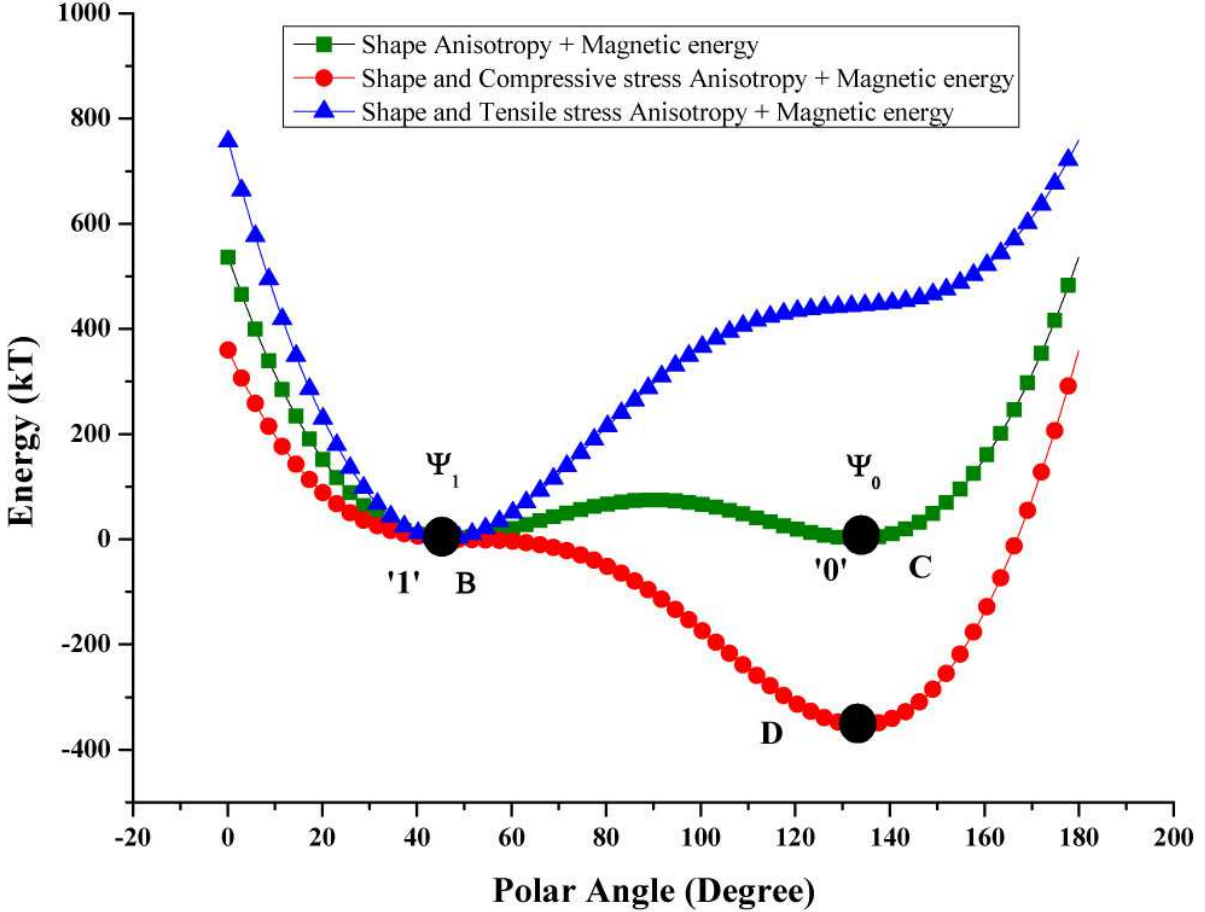}
\caption{{\bf Potential energy profiles of the magnetostrictive layer in Fig. 1
as a
function of its magnetization orientation}. Energy plot as a
function of polar angle ($\theta$) of the magnetization vector, where the green line is for the unstressed
magnet, the red line is for the compressively stressed magnet (-30 MPa),
and the blue line
is for the expansively stressed magnet (+37.5 MPa).
The voltage levels between $M$ and $N$
that generate these stresses are 56.3 mV are -70.375 mV respectively. }
\end{figure}

The gate works as follows: Let us first assume that the binary logic bits
`1' and `0' are encoded in voltage levels $V_0$ and $V_0/2$ [what determines the minimum value of
$V_0$ is discussed later]. The bias voltage is
set to $V_{BIAS}$ = $2V_0/3$. Every logic operation
is preceded by a RESET operation where the two inputs $V_{in1}$ and
$V_{in2}$ are set to $V_0/4$. During RESET, the potential drop appearing
between the terminals $M$ and $N$ in Fig. 1 is $V_{MN} = -5V_0/12$ [see supplementary material for a derivation],
which generates in-plane tensile stress in the direction of the line joining the two electrodes
and in-plane compressive stress in the direction perpendicular to the line joining the two electrodes
[we assume that the piezoelectric layer has been poled in the appropriate direction to make a negative
voltage drop generate these signs of the in-plane stresses].
This moves the system to point $B$ in the energy profile in Fig. 2 where the
magnetization vector is in state $\Psi_1$, nearly anti-parallel to the magnetization of the
top magnet (SAF). This makes
the resistance of the MTJ `high'. When the input voltages are subsequently
withdrawn by grounding the inputs and shorting the bias voltage source connected to the Si substrate, $V_{MN}$ drops to nearly zero. Therefore, the stress in the magnet relaxes,
but the system remains at point $B$.
Consequently, the MTJ is always
left in
the high resistance state after the RESET step is completed.

\begin{figure}
\includegraphics[width=6.5in]{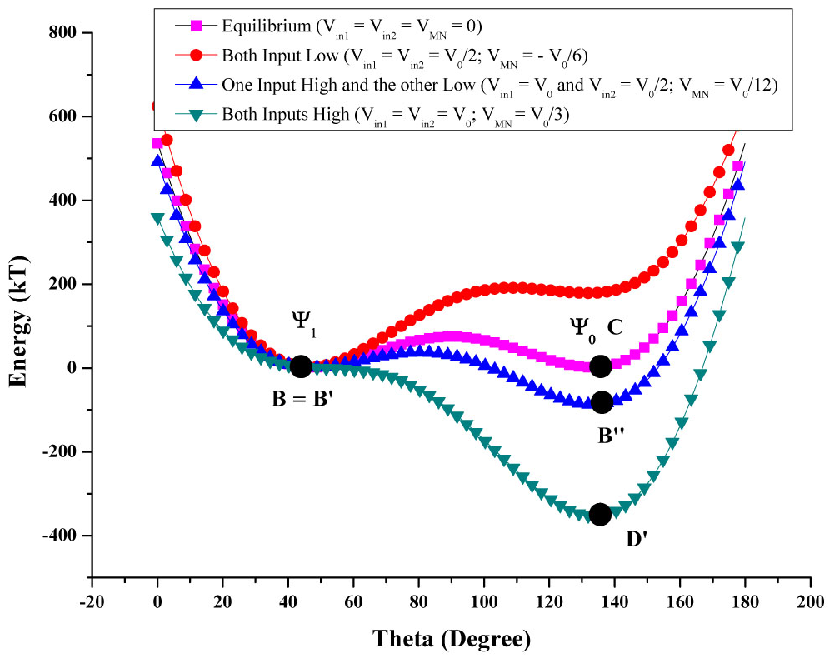}
\caption{{\bf Potential energy profiles of the magnetostrictive layer in Fig. 1
for different logic inputs}. Energy plot as a
function of polar angle ($\theta$) of the magnetization vector. The
RESET operation brings the magnetization to state $B$ where the magnetization is oriented along $\Psi_1$
and the MTJ resistance is high.
During logic operation, when both inputs are {\it low}, the magnet is under small tensile stress (+15 MPa)
and the global energy minimum
shifts
slightly to $B^{\prime}$ ($B \approx B^\prime$). Hence, the magnetization vector remains oriented
very close to $\Psi_1$ and the MTJ resistance remains high. If either input is {\it low},
 the magnet is under small compressive stress
(-7.5MPa) and the global energy minimum moves to B$^{\prime \prime}$. However, there is
an energy {\it barrier} of 36 kT separating $B$ and $B^{\prime \prime}$,
which cannot be transcended at room temperature. Consequently, the magnetization remains stuck at the local
minimum near $B$
 and
the MTJ resistance remains high. When both inputs are {\it high},
the magnet experiences high compressive stress (-30 MPa), which makes the energy profile
monostable with a single energy minimum at $D^\prime$ and no local minimum where the system
can be trapped. Therefore, the system migrates to $D^\prime$, the magnetization vector orients
close to $\Psi_0$, and the MTJ resistance goes low.
}
\end{figure}

In the logic operation stage, the following scenarios occur:
(1) if both inputs are low (i.e. $V_{in1} = V_{in2} = V_0/2$),
then $V_{MN} = -V_0/6$ [again, see the supplementary material for a derivation];
(2) if either input is low (i.e. $V_{in1} = V_0$ and $V_{in2} = V_0/2$, or vice versa), then
$V_{MN} = V_0/12$ (see supplementary material). The potential energy profiles for these
two scenarios are shown in Fig. 3. When both inputs are low, the global energy minimum is at $B' \approx B$.
Since the RESET operation left the system at $B$, the magnetization barely rotates
and the MTJ resistance remains high.
When one input is high and the other low, the {\it global} energy minimum moves to $B^{\prime \prime}$ which is closer
to the other stable magnetization orientation, but there is still a {\it local} energy minimum
close to $B$ which is separated from $B^{\prime\prime}$ by a sufficiently high potential barrier that {\it cannot be crossed}. Therefore, the
 system remains stuck in the metastable state corresponding to the local minimum near $B$ and the magnetization
does not rotate perceptibly. Hence, once again, the MTJ resistance remains high.
After the inputs are removed by grounding $V_{in1}$ and $V_{in2}$, shorting the bias voltage sources and
open-circuiting the bias current source, the strain in the
magnet relaxes and the magnetization  settles into the only accessible stable
state $B$. It remains there in perpetuity, thereby implementing {\it non-volatile} logic (memory
of the last output state is retained).
However, (3) if both inputs are high,
then $V_{MN} = +V_0/3$ (see supplementary material) and the strain becomes in-plane compressive
in the direction of the line joining the two electrodes and in-plane tensile in the direction perpendicular to the line joining the two electrodes.
This is sufficient to
change the potential energy
profile dramatically as shown in Fig. 3. Now the
operating point moves to $D'$ since it becomes the global minimum and there is no local minimum where the
system can get stuck.
Consequently, the magnetization vector rotates to an orientation nearly perpendicular to the magnetization of the
top
layer [state `$\Psi_0$' in Fig. 1(b)]. The resistance of the MTJ then drops
by $\sim$50\% since the resistance is inversely proportional to 1 + $\eta_1\eta_2$
$cos \gamma$,
where $\gamma$ is the angle between the magnetizations of the top (hard) and bottom (soft)
magnets. Since $\gamma$ is 180$^{\circ}$ and 90$^{\circ}$ in the high- and low-resistance 
states, the resistance ratio is $1/\left ( 1 - \eta_1 \eta_2 \right)$, which is
roughly 2:1, assuming that the spin injection and detection efficiencies of the magnet-spacer interfaces $\eta_1$ and $\eta_2$
are $\sim$70\% \cite{book} [if the efficiencies are less than 70\%, the logic levels will be encoded in $V_0$
and $xV_0$, where $x > 0.5$]. Subsequent removal of the input voltages (by grounding them),
drives the system to state $C$ where the MTJ resistance
remains low, thereby retaining memory of the last output state (non-volatility). The probability of the gate working
in this fashion, in the presence of thermal noise, has been calculated rigorously from stochastic Landau-Lifshitz-Gilbert
simulations of the magnetodynamics (see supplementary material) and that probability
was found to exceed 99.999999\% in all cases.

Let us now explain how this translates to NAND logic. Since there is not much electric field
in the PZT directly under the MTJ stack \cite{Lynch2013}, we can neglect any voltage drop in the PZT between
the magnetostrictive
magnet and the silicon substrate. Therefore, $V_{out} \approx I_{BIAS}R_{MTJ}$, where $R_{MTJ}$ is the resistance
of the MTJ stack.
The biasing constant current source $I_{BIAS}$ is set to $V_0/R_{high}$,
where $R_{high}$
is the resistance of the MTJ in the high-resistance state. Therefore, whenever
the MTJ is in the high resistance state, the output voltage is $V_0$ and whenever the
MTJ is in the low resistance state, the output voltage is $I_{BIAS} R_{low}$ = $V_0/2$
because $R_{low} = R_{high}/2$ [$R_{low}$ is the resistance of the MTJ in the low resistance state].
Since the logic bit 1 is encoded in voltage $V_0$ and logic bit 0 is encoded in the voltage
level $V_0/2$, we find that the output bit is 1 when either input bit
is 0, and it is 0 when both inputs are 1. In other words, we
have successfully implemented  a NAND gate (see the truth table shown in Fig. 1).

Let us now examine if this device fulfills all the requirements of a Boolean
logic gate.

\paragraph{Concatenability:} For concatenability,
the output
voltage of a preceding gate has to be fed directly to the input of a succeeding gate.
This requires that $V_{in1} (high)  = V_{in2} (high ) = I_{BIAS}R_{high} = V_0$,
and $V_{in1} (low)  = V_{in2} (low ) = I_{BIAS}R_{low} = V_0/2$
which is easily achieved by choosing $I_{BIAS} = V_0/R_{high}$. In the event the logic
levels have to be encoded in $V_0$ and $xV_0$ ($0.5 \le x \le 1$), the resistive network
at the input side and $V_{BIAS}$ have to be re-designed, but this is trivial.

\paragraph{Non-linearity:} Since the MTJ resistance has only
two values (high and low), the gate is inherently
 non-linear
\cite{behin-aein}.

\paragraph{Isolation between input and output:} The output and input terminals are completely isolated.
The input is provided to the two contact pads $E$ and $E'$ on the piezoelectric film that generates a strain in the
soft layer of the magneto-tunneling junction (MTJ) and changes the MTJ's resistance. This change in resistance causes the
output voltage $V_{out}$ to change. The output voltage is tapped from the top contact of the MTJ which is {\it electrically
isolated} from the input terminals. Therefore, this is effectively a {\it three-terminal} device -- input node, output node
and ground (common terminal) -- much like a transistor. The input and output nodes are electrically isolated.
Note that a change in the input voltage causes a change in the output voltage, but not the other way around.
Therefore, the input dictates the output, but not vice versa, resulting in input/output isolation.

There has been extensive discussions about input/output isolation in MTJ-based logic. Ref. [\onlinecite{Datta-review}]
discusses input and output states encoded in the magnetization states of two magnets via dipolar coupling,
while ref. [\onlinecite{m-logic}] discusses coupling via exchange between input magnetic domains and output. Both dipole and exchange
coupling are {\it bidirectional}; therefore, there is some chance that the output magnet's (or domain's) state
can influence the input magnet's (or domain's) state. This possibility does not exist in our case at all
since there is
no bidirectional coupling between input and output.

\paragraph{Gain:} Gain is ensured when the energy to switch the output
bit does not come
from the input energy, but from an independent power source
\cite{behin-aein}, which, in our case, is the
constant current source. Whenever the inputs $V_{in1}$ and $V_{in2}$ end up
switching the MTJ resistance,
the independent current source $I_{BIAS}$
switches $V_{out}$. The actual gain is the ratio of the swing in the output voltage to the swing in the
input voltage. In the supplementary material, we have calculated this gain
precisely and show it to be greater than 5 for the parameters we chose. The swing in the output voltage (and hence the
gain) can be increased by either increasing the bias current $I_{BIAS}$, or the MTJ-resistance $R_{MTJ}$.
The former approach will increase the energy dissipation and the latter will increase the charging time of the
input nodes -- both of which will ultimately increase the energy-delay product (see the
supplementary material). Thus, there is a trade-off
between gain and energy-delay product.

\paragraph{Universal logic:} The gate performs NAND operation which is universal.

\paragraph{Scalability:} Because we do not use magnetic fields to switch specific gates
(unlike refs. [\onlinecite{cowburn}] and [\onlinecite{ney}]),
but instead use only voltages, we do not have to space gates far apart so that
fringing magnetic fields from one gate do not influence the neighbor. As a result,
gates can be placed close to each other, thereby increasing the gate density. The gates can scale all
the way down to the superparamagnetic limit of the nanomagnets at the operating temperature.

\paragraph{Error-resilience:} Two types of errors afflict non-volatile
gate operation: {\it static} errors caused by the magnetization of the soft
magnetostrictive  layer
flipping spontaneously between its two stable orientations owing to thermal noise [thereby switching the output bit
erroneously in standby state],
and {\it dynamic} errors
that occur (also because of thermal noise) when the output switches to an
incorrect state in response to the
inputs changing. The static error probability is determined by the
energy barrier separating the
two stable magnetization
states in the soft layer. The minimum barrier
height is determined by the magnetic field strength,
the dimensions of the magnet and material parameters.
In our case, it was 73.1 kT at room temperature (see supplementary material),
so the static error
probability is $\sim e^{-73.1}$ $\approx$ 10$^{-32}$ per spontaneous switching attempt \cite{brown}.
In other words,
the retention time of an output bit in the
non-volatile logic gate at room temperature will be $\sim \left (1/f_0 \right)e^{73.1}$ =
1.77$\times$10$^{12}$ years, since the attempt frequency $f_0$ in nanomagnets will very rarely exceed 1 THz\cite{gaunt}.
In other words, the gate is indeed non-volatile.
Dynamic gate errors, however, are much more probable and accrue from two sources:
(1) thermal noise
causing erratic magnetization dynamics that drive magnets to the wrong stable
magnetization state resulting in bit error, and (2) complicated clocking schemes
that require precise timing
synchronization for gate operation and whose failure cause bit errors. The gate in
ref. [\onlinecite{behin-aein}] works with Bennett clocking \cite{bennett} which is predicated on the principle of
placing the output magnet in its
maximum energy state, and then waiting for the input signal
to drive it to the desired one among
its two minimum energy states to produce the correct output bit.
This strategy is risky
 since the maximum energy state is also maximally unstable. While perched
 on the energy maximum, thermal fluctuations
can  drive the output magnet to the
wrong minimum energy state with unacceptably high probability
\cite{roychowdhury}, resulting in unacceptable bit error rates. A later modification \cite{behin-aein2}
overcame this shortcoming, but at the expense of much increased
energy dissipation. Moreover, that logic gate also
requires a complicated
clocking sequence without which it cannot operate. In contrast,
we {\it never} place any element of our gate
at the maximum energy state (no Bennett clocking) and no complicated clocking sequence is needed.

An important consideration for Boolean logic is {\it logic level restoration} \cite{jackson}. If noise broadens the
input voltage levels $V_0$ and $V_0/2$, making it harder to distinguish between bits 0 and 1, the logic device should
be able to
restore the distinguishability by ensuring that the output voltage levels are not broadened and remain well separated.
For this to happen, the transfer characteristic of the gate (when used as an inverter) must show a sharp transition.
We have computed the transfer characteristic ($V_{out}$ versus $V_{in}$) by shorting the two inputs (thus making it an inverter) and calculating
the output $V_{out}$ for various values of $V_{in}$ at room temperature in the presence of thermal noise. The calculation procedure is described in the supplementary
section. The characteristic is shown in Fig. 4 and the sharpness of the transition allows for excellent logic level 
restoration capability.

\begin{figure}
\includegraphics[width=6.5in]{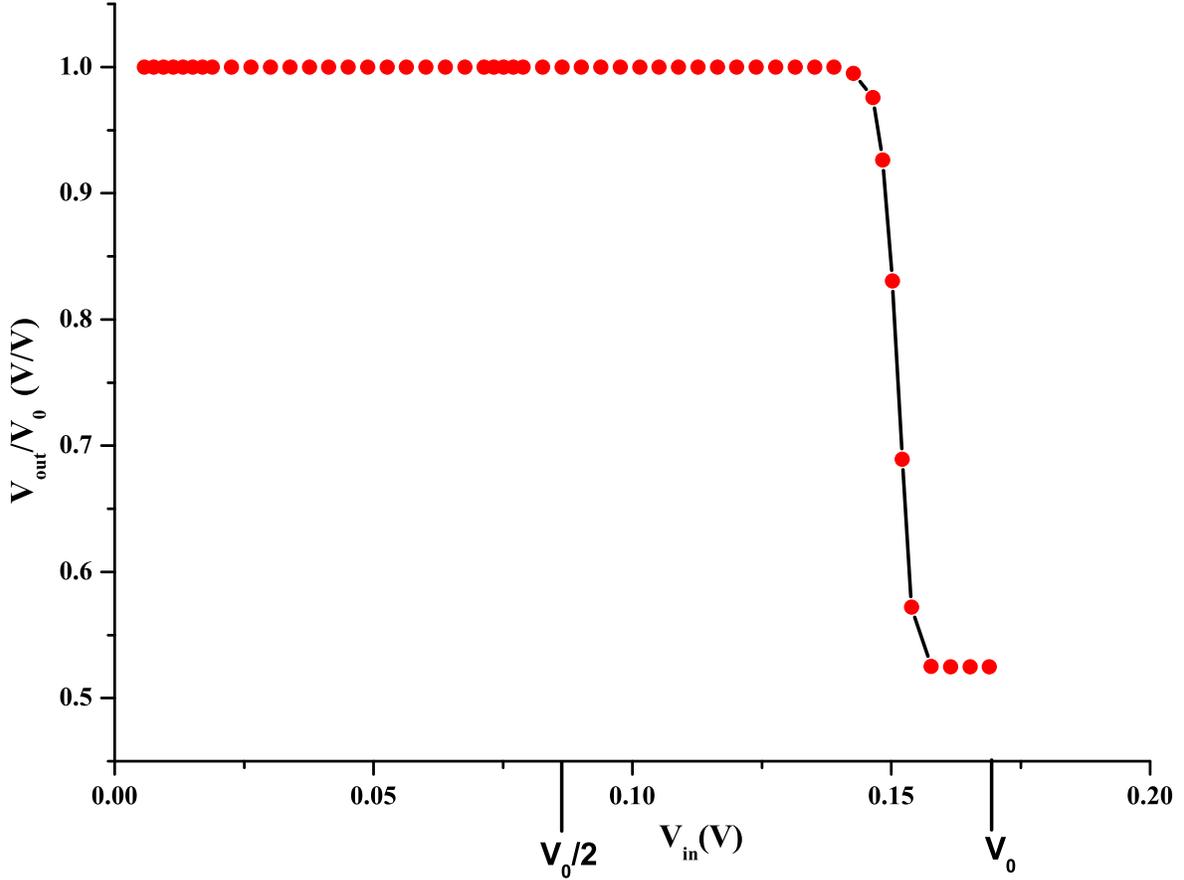}
\caption{{\bf Transfer characteristic in the inverter mode}. Shorting the two inputs of a NAND gate makes it an inverter.
Plot of $V_{out}$ versus $V_{in}$ of the inverter at room temperature, where the $V_{out}$ values have been thermally averaged.
}
\end{figure}

The proposed gate has unprecedented energy-efficiency
that far exceeds that of other non-volatile magnetic NAND gates. There are two contributions to the energy
dissipated in this logic gate during a logic operation: direct dissipation associated with switching the
gate (which has two components --
internal dissipation due to Gilbert damping
that occurs in the magnet during magnetization rotation, and  $C \left ( V_{MN}
\right )^2$ energy
dissipated in turning on/off the potential $V_{MN} = V_0/3$ (= 56.3 mV) abruptly or
non-adiabatically with  $C$ being the capacitance between
the shorted pair of electrodes and the n$^+$-Si substrate), and the indirect dissipation  in the
resistors $R$, $R_P$ and $R_{MTJ}$ due to the bias current source $I_{BIAS}$.
These contributions are computed in the supplementary
section and add up to a mere 5176 kT (21.44 aJ) at room temperature. This dissipation is
at least an order of magnitude less than in a CMOS transistor embedded in a circuit \cite{Datta-review}.
The switching time, on the other hand,
is $\sim$1.3 ns, which is almost
one order of magnitude
longer than that of the CMOS based logic gate. However, the CMOS based gate
is volatile while this gate is non-volatile.
The time-averaged energy delay product of this gate is 2.78$\times$10$^{-26}$ J-s, which is about
two orders of magnitude superior to that of any other magnetic (non-volatile) logic gate
and one order of magnitude superior to CMOS. The gate error probability, on the other hand, is
10$^{-8}$ per logic operation.

Logic gates of this type may have a special niche for medically implanted processors
such as pacemakers \cite{geddes}, wearable electronics \cite{starner} or devices implanted in an epileptic patient's
brain that monitor
 brain signals
and warn of an impending seizure. They need not be very accurate and therefore a gate error
probability of 10$^{-8}$ may be tolerable. Most importantly, they will dissipate so little
energy that they can be powered by the user's body movements alone and not require
a battery \cite{kuntal-apl}.

\section*{Methods}

{\small To fabricate the
gate, a piezoelectric (PZT) thin film ($\sim$50 nm thick) is deposited on a conducting n$^+$-Silicon substrate
which is grounded through a bias voltage $V_{BIAS}$. A skewed MTJ stack is fabricated on top of the PZT film.
 The bottom layer material is chosen as Terfenol-D because of its large magnetostriction (600 ppm).
The magnetostriction is positive which tends to make the magnetization align along the direction of tensile stress
and perpendicular to the direction of compressive stress.
The angle between the
major axes of the two elliptical nanomagnets is determined by the angular separation between $\Psi_1$ and $\Psi_0$. The current source
$I_{BIAS}$
is connected across the MTJ stack. The
magnetostrictive nanomagnet has a major axis of 100 nm, minor axis of 42 nm and thickness of 16.5 nm,
which ensures
that it has a single ferromagnetic domain.

To evaluate the dynamic error probability, the magnetization dynamics of the soft magnetostrictive
magnet induced by stress in the presence of thermal
noise is modeled by the stochastic Landau-Lifshitz-Gilbert equation
\cite{roychowdhury}. In the Supplementary section, we present
results of  simulations
to show that if $V_0$ = 0.1689 V, then switching is
accomplished in 1.3 ns and the dynamic error probability associated with incorrect switching is less than 10$^{-8}$ in
every gate operation if we keep the voltage on for 1.3 ns.
Therefore, the gate can work at a clock frequency of $\sim$1/1.3 ns $ > $ 0.75 GHz with an error probability $<10^{-8}$.
Stated succinctly, the probability of the output voltage being low when both inputs
are high is $>99.999999$\% and the probability of it being low when either input is
low is $<10^{-8}$. In other words, the NAND gate works with $>99.999999$\% fidelity.
This is unimpressive for
transistor-based volatile logic, but it is
remarkable for non-volatile magnetic logic gates, which typically
have very high error probabilities \cite{salehi,roychowdhury,bokor}.
This degree of error-resilience may be sufficient for use in stochastic logic architectures \cite{Qian}.}

\newpage

\begin{center}
{\Large \bf SUPPLEMENTARY MATERIAL}
\end{center}

In this accompanying supplementary material, we elucidate gate operation, concatenation and choice of the voltage level $V_0$.
We also describe the stochastic Landau-Lifshitz-Gilbert simulations, and calculations of the transfer characteristic
and energy dissipation in a gate operation.

\section{Gate operation}
To understand how the RESET, logic and the concatenation schemes work, consider Fig. 5. The nodes
$M$ and $N$ represent the same nodes as in Fig. 1(a) of the main paper and $V_{MN}$ is
the voltage drop between these nodes. Therefore, $V_{MN}$ is the voltage drop across the piezoelectric layer
that generates strain in the magnetostrictive layer (soft layer of the MTJ) and makes its magnetization rotate.
Note that $V_{MN}$ alone determines the MTJ resistance. As established by the energy profiles in the main paper,
when $V_{MN}$ is either negative or slightly positive, the hard and soft layers of the MTJ remain magnetized
in anti-parallel directions and the MTJ resistance remains high. This high resistance is denoted by $R_{high} = R_0$.
When $V_{MN}$ is positive and sufficiently large in magnitude, the magnetizations of the hard and soft layers
become mutually perpendicular and the MTJ resistance drops by a factor of 2 to become $R_{low} = R_0/2$.

Since the ratio $R_{high}/R_{low}$ is 2:1, logic `1' must be encoded in some voltage level $V_0$ and logic `0' in
voltage level $V_0/2$. This is needed because the logic levels at the output are determined solely
by the MTJ resistance.

\begin{figure}
\includegraphics[width=6.5in]{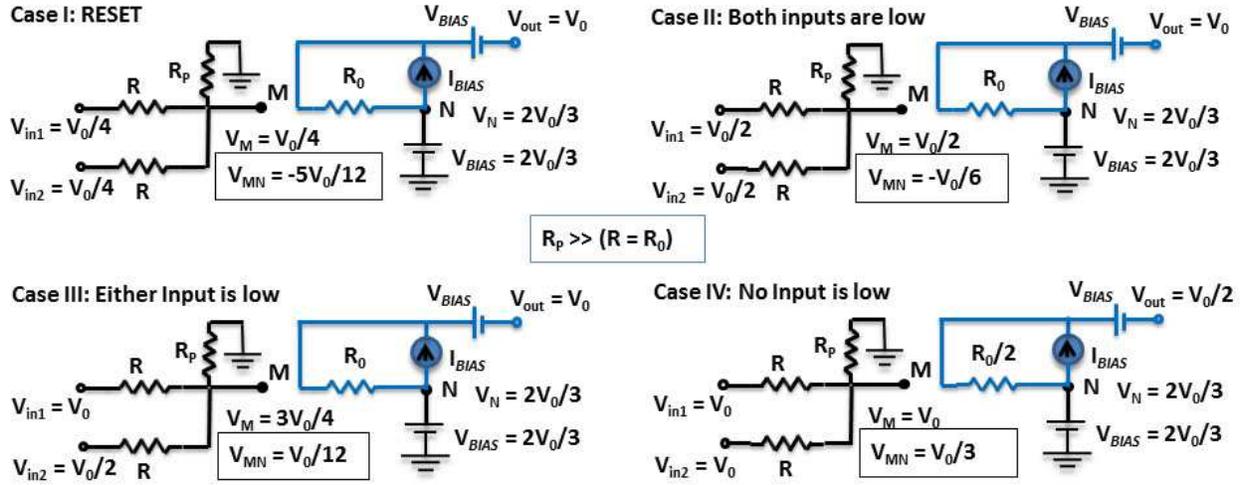} \label{fig:LogicOP}
\caption{{\bf  Logic operations}}
\end{figure}

Let us consider the RESET operation that is supposed to leave the MTJ resistance in the high state $R_0$
(Case I in Fig. 5). The input voltages are set to V$_0$/4. The voltage at node $M$ is then found
by superposition and it is V$_0$/4. Since the bias voltage V$_{BIAS}$ is set to 2V$_0$/3, the voltage at node $N$
is always fixed at 2V$_0$/3. Therefore, $V_{MN}$, which is the voltage drop across the PZT thin film, becomes
V$_0$/4 - 2V$_0$/3 = -5V$_0$/12. This negative voltage generates a stress profile
in the magnetostrictive layer that leaves its
magnetization pointing anti-parallel to that of the the hard (SAF) layer of the MTJ (close to $\Psi_1$).
Therefore, the MTJ resistance $R_{MTJ}$ is left high at $R_{high} = R_0$ by the RESET step.

Note that the voltage drop between the bottom (soft) layer of the MTJ and node $N$ is almost
zero since the metallic layer shorts out the electric field underneath it in the PZT \cite{Lynch2013}.
Therefore, by applying Kirchoff's voltage law in the output loop, we find that
\begin{equation}
V_{out} \approx -V_{BIAS} + I_{BIAS}R_{MTJ} +V_{BIAS} \approx  V_0{{R_{MTJ}}\over{R_0}},
\label{output}
\end{equation}
since $I_{BIAS}$ is set to $V_0/R_0$. Because $R_{MTJ} = R_{high} = R_0$ after the RESET stage, $V_{out} = V_0$
after the RESET operation has been completed.

Next consider the logic operation stage when both inputs are low (Case II). Since
$V_{in1}$ = $V_{in2}$ = $V_0/2$, $V_M$ is $V_0/2$ [use superposition] and thus $V_{MN}$ (= $V_M - V_{BIAS}$) is -$V_0/6$.
This negative voltage once again generates a stress profile in the magnetostrictive magnet that leaves the
magnetizations of the hard and soft layers of the MTJ anti-parallel and the MTJ resistance high.
Therefore, from Equation (\ref{output}), $V_{out} = V_0$. In other words, when both inputs are `0', the output
is `1'.

When one input is high and the other low (Case III), $V_{in1}$ = $V_0/2$ and $V_{in2}$ = $V_0$ (the case where
$V_{in1}$ = $V_0$ and $V_{in2}$ = $V_0/2$ is completely equivalent). The voltage at node $M$,
 $V_M$, is now $3V_0/4$ [again, by superposition], which makes $V_{MN} = V_0/12$. The stress profile
 in the magnetostrictive layer now changes, but not enough to rotate its magnetization by overcoming the shape anisotropy energy barrier of the
elliptical magnet.
Therefore, MTJ resistance remains high at $R_0$ and the output voltage remains high at $V_0$. In
other words,  when one input is `1' and the other is `0', the output is `1'.

When both inputs are high (Case IV), $V_{in1}$ = $V_{in2}$ = $V_0$. In that case, V$_{M}$ changes to
$V_0$ [use superposition], and $V_{MN}$ becomes $V_0$/3. The stress generated by this magnitude of $V_{MN}$ in the 
magnetostrictive layer of the MTJ
is sufficient to overcome the shape anisotropy barrier.
As a result, the magnetization of the soft layer of the MTJ
now rotates by $\sim$90$^{\circ}$, placing it approximately perpendicular to that of the hard layer. Therefore,
the MTJ's resistance drops to $R_0/2$  and [from Equation (\ref{output})] the output voltage drops to  $V_0/2$.
Thus, when both inputs are `1', the output is `0'. All this implements NAND logic.

Note that the output voltage levels encoding bits `1' and `0' are $V_0$ and $V_0/2$ which are also the input voltage
levels encoding bits `1' and `0'. Therefore, the output of one stage can be directly fed to the next stage as input
without requiring additional hardware for amplification or level shifting. That makes this construct {\it concatenable}.

\section{Choice of voltage level $V_0$}

In order to choose the value of $V_0$ (which ultimately determines the amount of dissipation, switching delay and
energy-delay product), we have to ensure that compressive stress generated by $V_{MN} = V_0/3$
is sufficient to overcome the shape anisotropy barrier in the elliptical magnetostrictive layer and rotate its
magnetization, but compressive stress generated by $V_{MN} = V_0/12$ is not. The amount of stress generated by a
certain voltage, and the effective shape anisotropy barrier in the presence of the permanent magnetic field,
 depend on many parameters such as the strength of the magnetic field, the shape and size of the magnetostrictive layer,
the electrode size and placements, the piezoelectric layer thickness, and the piezoelectric and magnetostrictive
materials. For the choices we made, we found from stochastic Landau-Lifshitz-Gilbert simulations  of magnetodynamics
in the presence of
room-temperature thermal noise \cite{Roy2012} that a compressive stress of -30 MPa rotates the magnetization
with greater than 99.999999\% probability (and switches the MTJ resistance from high to low) in the
presence of room-temperature thermal fluctuations, while a compressive stress of -7.5 MPa
(one-fourth of -30 MPa) has less than 10$^{-8}$ probability of rotating the magnetization and switching the MTJ resistance.
Therefore, $V_{MN} = V_0/3$ needs to generate a stress of
-30 MPa (compressive strain is negative).
The material chosen for the magnetostrictive material is Terfenol-D because of its large magnetostriction.
From the Young's modulus of Terfenol-D, we calculated that the strain required to generate a stress
of -30 MPa is -3.75$\times$10$^{-4}$. To generate this amount of strain,
the strength of the electric field in the PZT between the shorted electrodes and the n$^+$-Si substrate should be
1.126 MV/m (interpolated from the results in
 Ref. [\onlinecite{Lynch2013}]). This value is well below the breakdown field of PZT. We assume that the PZT layer thickness is 50 nm
 (easily achievable); hence,
the voltage $V_{MN}$ needed to generate the strain of -3.75$\times$10$^{-4}$ will  be 1.126 MV/m $\times$ 50 nm =
56.3 mV. Hence,
$V_0 = 3V_{MN}$ = 0.1689 V.

\section{Stochastic Landau-Lifshitz-Gilbert (LLG) simulations}

The error probability associated with gate operation, the internal energy dissipated during switching, and the switching
delay -- all in the presence of room-temperature thermal noise -- are calculated from the stochastic
Landau-Lifshitz-Gilbert (LLG) equation$\footnote{The LLG equation is based on the macrospin assumption which assumes 
that the entire domain in a single-domain nanomagnet rotates coherently and all the spins in it rotate in unison,
collectively acting as one giant classical spin. In general, the 
macrospin assumption works fairly well for elliptical shaped nanomagnets and this has been verified by comparison with micromagnetic 
simulations (OOMMF) \cite{mamun}.}$.
 We first write expressions for the various contributions to the potential energy of the magnetostrictive layer
and then find the effective torques acting on the magnetization vector due to these contributions as well as the random torque due to thermal noise. These
torques rotate the magnetization vector. The entire procedure is
described next.

\begin{figure}
\centering
\includegraphics[width=3.4in]{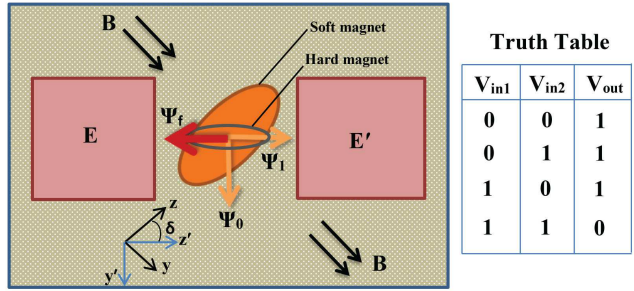} \label{NAND_gate}
\caption{{\bf  The fixed magnetization
orientation of the top (hard) magnet is denoted by $\Psi_f$, and the two stable magnetization orientations
of the bottom (soft) magnet are denoted by $\Psi_0$ and $\Psi_1$. The MTJ resistance is high when the soft magnet's
magnetization is aligned along $\Psi_1$. The MTJ resistance is (ideally)
a factor of
2 lower when the soft magnet's
magnetization is aligned along $\Psi_0$. The slanted ellipse is the footprint
of the soft magnet
and the horizontal ellipse is the footprint of the hard magnet. The black double arrows show the direction of
the permanent magnetic field.}}
\end{figure}

We reproduce Fig. 1(b) from the main paper as Fig. 6 and define our coordinate system such that the magnet's easy (major) axis
lies along the z-axis and the in-plane hard (minor) axis lies along the
y-axis (see also Fig. 1(a) in main paper). Application of a positive/negative voltage between the electrode pair $EE'$
and the conducting n$^+$-Si substrate generates biaxial strain leading to compression/expansion
along the z$^\prime$-axis and expansion/compression along the y$^\prime$-axis \cite{Lynch2013}.
The latter two axes are the axes of $\Psi_1$ and $\Psi_0$. The angle between the z- and z$^\prime$ axes is $\delta$, 
which is therefore the angle between the major axes of the hard and soft elliptical magnets.

To derive general expressions for the
instantaneous potential energies of the nanomagnet due to shape-anisotropy, stress-anisotropy and the static magnetic
field, we used the primed axes of reference (x$^\prime$, y$^\prime$, z$^\prime$) and represented the magnetization orientation of the
single-domain magnetostrictive magnet in spherical coordinates with
$\theta^{\prime}$ representing the polar angle and $\phi^{\prime}$ representing the azimuthal angle.
The magnitude of the magnetization is invariant in time and space owing to the macrospin assumption.

Using  the rotated coordinate system (see Fig. 6), the shape anisotropy energy of the nanomagnet
$E_{sh}(t)$ can be written as,
\begin{eqnarray}
E_{sh}(t) & = & E_{s1}(t) {\sin}^2{\theta^{\prime}(t)} + E_{s2}(t) {\sin 2\theta^{\prime}(t)} \nonumber\\
          & + & \frac{\mu_{0}}{2} \Omega {M}^2_{s} ( N_{d-yy}{\sin}^2{\delta} + N_{d-zz} {\cos}^2{\delta})\nonumber\\
E_{s1}(t) & = & \left( \frac{\mu_{0}}{2} \right) \Omega {M}^2_{s} \{ N_{d-xx}{\cos}^2{\phi^{\prime}(t)} + N_{d-yy}{\sin}^2{\phi^{\prime}(t)} {\cos}^2{\delta} \nonumber\\
          & - & N_{d-yy} {\sin}^2{\delta}  + N_{d-zz}{\sin}^2{\phi^{\prime}(t)} {\sin}^2{\delta}  - N_{d-zz} {\cos}^2{\delta} \} \nonumber\\
E_{s2}(t) & = & \left( \frac{\mu_{0}}{4} \right) \Omega {M}^2_{s} \left( N_{d-zz} -  N_{d-yy} \right){\sin \phi^{\prime}(t)} {\sin 2\delta},
\end{eqnarray}
where $\theta^{\prime}(t)$ and $\phi^{\prime}(t)$ are respectively the instantaneous polar and azimuthal angles of the
magnetization vector in the
rotated frame,
$M_s$ is the saturation magnetization of the magnet, $N_{d-xx}$, $N_{d-yy}$ and $N_{d-zz}$ are the demagnetization factors
that can be evaluated from the nanomagnet's dimensions \cite{Beleggia2005}, $\mu_0$ is the permeability of free space,
and $\Omega = (\pi/4)abd$ is the nanomagnet's volume.

The potential energy due to the static magnetic flux density $B$ applied along the in-plane hard axis is given by
\begin{equation}
E_{m}(t) =  M_s \Omega B \left [ {\cos \theta^{\prime}(t)} {\sin \delta} - {\sin \theta^{\prime}(t)} {\sin \phi^{\prime}(t)} {\cos \delta} \right ].
\end{equation}

 The stress anisotropy energy is given by
\begin{equation}
E_{str}(t) = - \frac{3}{2}\lambda_{s} \epsilon(t) Y \Omega {\cos}^2{\theta^{\prime}(t)},
\end{equation}
where $\lambda_s$ is the magnetostriction coefficient, $Y$ is the Young's modulus, and $\epsilon(t)$ is the strain
generated by the
applied voltage $V_{MN}$ at the instant of time $t$. We only consider the uniaxial strain along the line joining
the two electrodes, but the strain is actually biaxial resulting in tension/compression along that line and compression/tension
along the perpendicular direction. The torques due to these two components {\it add}.
 Therefore, we {\it underestimate} the stress anisotropy energy, which makes
all our figures {\it conservative}.

We neglect any contribution due to the dipolar interaction of the hard magnet since the use of the synthetic
anti-ferromagnet makes it negligible.

The total potential energy of the nanomagnet at any instant of time $t$ is therefore
\begin{equation}
E(t) = E \left ( \theta^{\prime}(t), \phi^{\prime}(t) \right ) = E_{sh}(t) + E_{m}(t) + E_{str}(t).
\end{equation}
The above result is used to plot the energy profiles in the main paper as a function of $\theta$
for $\phi = 90^{\circ}$ under various scenarios.

We follow the standard procedure to derive the time evolution of the polar and azimuthal angles of the magnetization
vector
in the rotated coordinate frame under the actions of the torques due to shape anisotropy, stress anisotropy, magnetic field
and thermal noise.

The torque that rotates the magnetization of the shape-anisotropic magnet in the presence of stress can be written as
\begin{eqnarray}
\mathbf{\tau_{ss}}(t)  & = &  - \mathbf{m}(t) \times \left(\frac{\partial E}{\partial \theta^{\prime}(t)} \hat{\boldsymbol{\theta}} + \frac{1}{\sin \theta^{\prime}(t)}
\frac{\partial E}{\partial \phi^{\prime}(t)} \hat{\boldsymbol{\phi}}\right) \nonumber\\
                      & = & \{E_{\phi 1}(t) \sin\theta^{\prime}(t) + E_{\phi 2}(t) \cos\theta^{\prime}(t) \nonumber\\
                      & - & M_s \Omega B \cos\delta \cos\phi^{\prime}(t)\}\hat{\boldsymbol{\theta}} \nonumber\\
                      & - & \{E_{s1} (t) \sin2\theta^{\prime}(t) + 2E_{s2} (t) \cos2\theta^{\prime}(t) \nonumber\\
                      & - &  M_s \Omega B (\cos\delta \sin\phi^{\prime}(t) \cos\theta^{\prime}(t) + \sin\delta \sin\theta^{\prime}(t))  \nonumber\\
                      & + & (3/2) \lambda_s \epsilon(t) Y \Omega \sin2\theta^{\prime}(t) \}\hat{\boldsymbol{\phi}},
\end{eqnarray}
where $\mathbf{m} (t)$ is the normalized magnetization vector, quantities with carets are unit vectors in the
original frame of reference, and
\begin{eqnarray}
E_{\phi 1}(t) & = & \frac{\mu_{0}}{2} {M}^2_{s} \Omega \{ \left( N_{d-yy} {\cos}^2{\delta}  + N_{d-zz} {\sin}^2{\delta} \right) \sin 2\phi^{\prime}(t)
               -  N_{d-xx}\sin 2\phi^{\prime}(t) \} \nonumber \\
E_{\phi 2}(t) & = & \frac{\mu_{0}}{2} {M}^2_{s} \Omega \left( N_{d-zz} - N_{d-yy} \right) \sin 2\delta \cos \phi^{\prime}(t). \nonumber
\end{eqnarray}

At non-zero temperatures,  thermal noise generates a random magnetic field $\mathbf{h} (t)$ with Cartesian
components $\left ( h_x(t), h_y(t), h_z(t) \right )$
that produces
a random thermal torque which can be expressed as \cite{Roy2012}
\begin{equation}
\mathbf{\tau_{th}}(t)\! =\! \mu_0 M_s \Omega \mathbf{m}(t)\! \times\! \mathbf{h}(t)\! =\! - \! \mu_0 M_s \Omega \left [ h_{\phi}(t) \hat{\boldsymbol{\theta}} -  h_{\theta}(t) \hat{\boldsymbol{\phi}} \right ],\nonumber
\end{equation}
where
\begin{eqnarray}
h_{\theta}(t) & = & h_x (t) \mbox{cos}\theta^{\prime}(t) \cos\phi^{\prime}(t) +  h_y (t) cos\theta^{\prime}(t) \sin\phi^{\prime}(t)
               -  h_z (t) sin\theta^{\prime}(t) \nonumber\\
h_{\phi}(t)   & = & -h_x (t) \sin\phi^{\prime}(t) +  h_y (t) cos\phi^{\prime}(t).
\end{eqnarray}

In order to find the temporal evolution of the magnetization vector under the vector sum of the different torques
mentioned above,
we solve the stochastic Landau-Lifshitz-Gilbert (LLG) equation:
\begin{eqnarray}
\frac{d\mathbf{m}(t)}{dt} & - &  \alpha \left[ \mathbf{m}(t) \times  \frac{d\mathbf{m}(t)}{dt} \right]
                           =  \frac{-|\gamma|}{\mu_0 M_s \Omega} \left ( \mathbf{\tau_{ss}}(t)+\mathbf{\tau_{th}}(t) \right )
\end{eqnarray}
From the above equation, we can derive two coupled equations for the temporal evolution of the
polar and azimuthal angles of the magnetization vector:
\begin{eqnarray}
\frac{d\theta^{\prime}(t)}{dt} & = & -\frac{|\gamma|}{(1+\alpha^2)\mu_0 M_s \Omega} \{ E_{\phi 1}(t) \sin\theta^{\prime}(t) + E_{\phi 2}(t) \cos\theta^{\prime}(t) \nonumber\\
                   & - &  M_s \Omega B \cos\delta \cos\phi^{\prime}(t) - \mu_0 M_s \Omega h_{\phi}(t) \nonumber\\
                   & + & \alpha \{ E_{s1} (t) \sin2\theta^{\prime}(t) - \mu_0 M_s \Omega h_{\theta}(t) \nonumber\\
                   & + & 2E_{s2} (t) \cos2\theta^{\prime}(t) + (3/2) \lambda_s \epsilon(t) Y \Omega \sin2\theta^{\prime}(t) \nonumber\\
                   & - &  M_s \Omega B (\cos\delta \sin\phi^{\prime}(t) \cos\theta^{\prime}(t) + \sin\theta^{\prime}(t) \sin\delta ) \} \}
                   \label{eq:finaltheta} \\
\frac{d\phi^{\prime}(t)}{dt}   & = & \frac{|\gamma|}{\sin\theta^{\prime}(t)(1+\alpha^2)\mu_0 M_s \Omega} \{ E_{s1} (t) \sin2\theta^{\prime}(t) \nonumber\\
                   & + & 2E_{s2} (t) \cos2\theta^{\prime}(t) + (3/2) \lambda_s \epsilon(t) Y \Omega \sin2\theta^{\prime}(t) \nonumber\\
                   & - &  M_s \Omega B (\cos\delta \sin\phi^{\prime}(t) \cos\theta^{\prime}(t) + \sin\delta \sin\theta^{\prime}(t)) \nonumber\\
                   & - & \mu_0 M_s \Omega h_{\theta}(t) - \alpha ( E_{\phi 1}(t) \sin\theta^{\prime}(t) + E_{\phi 2}(t) \cos\theta^{\prime}(t) \nonumber\\
                   & - &  M_s \Omega B \cos\delta \cos\phi^{\prime}(t) - \mu_0 M_s \Omega h_{\phi}(t) ) \}. \label{eq:finalphi}
\end{eqnarray}
Solutions of these two equations yield the magnetization orientation $\left ( \theta^{\prime}(t), \phi^{\prime}(t)
\right )$ at any instant of time $t$. Since the thermal torque is random, the solution procedure involves generating
switching trajectories by starting each trajectory with
an initial value of ($\theta^{\prime}, \phi^{\prime}$) and finding the values of these angles at any other time
by running a simulation using a
time step of $\Delta$t = 1 ps and for a sufficiently long duration. At each time step, the random thermal
torque is generated stochastically. The time step is equal to the inverse of
the maximum attempt frequency of demagnetization due to thermal noise in nanomagnets \cite{gaunt}. The duration of the
simulation is always sufficiently long to ensure that the final results are independent of this duration, and they are
also verified to be independent of the time step.

The permanent magnetic field ($B$ = 0.14 T) applied along the +y- direction (hard axis of the magnet)
 makes the two stable states of the soft magnet's magnetization align along $\Psi_1$ ($\theta = \theta_1$ = 46.9$^\circ$)
and
 $\Psi_0$ ($\theta = \theta_0$ = 133.1$^\circ$) leaving a separation angle $\gamma$ of 86.2$^\circ$
(Fig. 6) between them. Thermal noise however will make the magnetization of the soft magnet
fluctuate around either of
these two orientations and in order to determine the thermal distribution around $\Psi_1$ (which is
where the RESET operation leaves the magnetization at), we solve the last two equations
in the absence of any stress
by starting with the initial state $\theta$ = 46.9$^\circ$ and $\phi$ = 90$^\circ$
and obtaining the final values of $\theta$ and $\phi$ by running the simulation for a
long time. This process is repeated 100 million times. A histogram is then generated
from these 100 million trials for the final values of $\theta$ and $\phi$, which yields
the thermal distribution around $\Psi_1$.

To study the switching dynamics under the influence of stress,
we generate 100 million switching trajectories in the stressed state of the magnet by solving Equations
(\ref{eq:finaltheta}) and (\ref{eq:finalphi}), again using a time step of 1 ps. This time the initial
magnetization orientation for each of the 10$^8$ trajectories is chosen from the thermal
distributions generated in the previous step with the appropriate weightage
since the RESET step always leaves the magnetization around state $\Psi_1$.
The simulation is continued for 1.5 ns. We find that when the stress is either tensile
(+15 MPa corresponding to $V_{MN} = -V_0/6$), or compressive but weak (-7.5 MPa corresponding to $V_{MN} = V_0/12$),
the magnetization's polar angle returns to within 4$^{\circ}$ of $\Psi_1$ ($\theta = \theta_1$ = 46.9$^\circ$) in 1.3 ns
or less for every one of the 10$^8$ trajectories. After 1.3 ns, the stress is removed abruptly and the simulation
is continued for an additional 0.2 ns to ensure that the final state does not change. It did not change for any of
the 10$^8$ trajectories.  This procedure tells us that when the inputs to the logic gate are both low, or
one is high and the other is low, the magnetization of the soft layer of the MTJ does not rotate and the MTJ
resistance remains high with $>$99.999999$\%$ probability. This fulfills  the requirements of the NAND gate
with $>$99.999999$\%$ probability.

In the case of one input low and one input high, what prevents rotation from $\Psi_1$ to $\Psi_0$
is the energy barrier of 36 kT between
these two states as discussed in the main paper. This barrier is high enough to reduce the 
unintended switching probability to below 10$^{-8}$.

When both inputs are high, a compressive stress of -30 MPa is generated in the magnetostrictive magnet. Once again,
we pick the initial orientations of the magnetization from the thermal distribution around $\Psi_1$ which is where
the RESET step leaves the magnet at, and generate 100 million switching
trajectories
as before. This time $\theta$ approaches within 4$^\circ$ of final state $\Psi_0$ ($\theta = \theta_0$ = 133.1$^\circ$)
in 1.3 ns or less. We continue the simulation for an additional 0.2 ns to confirm that once the magnetization reaches
the vicinity of $\Psi_0$, it settles around that orientation and does not return to the neighborhood of the
initial orientation $\Psi_1$.
We repeated this procedure for 10$^8$ times and found that every single switching trajectory behaved in the above manner.
Therefore, we conclude  that when the inputs to the logic gate are both high,  the magnetization of the soft layer of
the MTJ
does rotate and the MTJ resistance goes low with $>$99.999999$\%$ probability. That fulfills  the remaining requirement of
the NAND gate with $>$99.999999$\%$ probability.

\section{Transfer characteristics of NAND gate}
When the two inputs of a NAND gate are shorted, it behaves like a NOT gate (inverter).
If the input voltage to the inverter is between $V_0/12$ and $V_0/3$ (i.e. the compressive stress is between
-7.5 MPa and -30 MPa), the energy profile becomes such that the magnetization of the soft layer
may rotate to an intermediate state between $\Psi_0$ and $\Psi_1$ and
fluctuate around that orientation because of thermal noise, as long as the stress is kept on. We can time-average over the fluctuations to
determine the `steady-state' mean
orientation at that input voltage $V_{in}$
and thence calculate $R_{MTJ}$ and $V_{out} = V_0R_{MTJ}/R_0$. In order to do this, we calculate the
stress generated by the $V_{MN}$ corresponding to the input voltage $V_{in}$ and run the stochastic Landau-Lifshitz-Gilbert
simulation to determine the mean steady-state magnetization orientation of the soft layer and from
that the mean values of $R_{MTJ}$ and $V_{out}$. The purpose of this exercise
is to find the transfer characteristic $V_{out}$ versus $V_{in}$ at room temperature. This characteristic is shown in the main paper and
shows a sharp transition. The transition range is from 142.69 mV to 157.7 mV (a range of 15.01 mV) of input voltage whereas the logic
levels are 88.67 mV and 168.9 mV. This portends excellent
logic level restoration capability. In the high state, the input voltage can drift down by 11.2 millivolts and still
produce the correct output state, while in the low state, the input voltage can drift up by 54.02 millivolts and
still produce the correct output state.

One can also calculate an effective ``gain'' from the transfer characteristic. We define the gain as
$\gamma = \Delta V_{out}/\Delta V_{in}$ and this quantity is (168.9 - 88.67)/15.01 = 5.34.

\section{Energy Dissipation}
The energy dissipated in the NAND gate has two components: direct (associated with the switching action)
and indirect (caused by the peripherals unavoidably).
 The direct dissipation has two contributions:
internal dissipation due to Gilbert damping that occurs while the magnetostrictive layer's magnetization
switches (rotates) and energy $C \left ( V_{MN}
\right )^2$ dissipated in turning on/off the potential $V_{MN}$  abruptly or
non-adiabatically during the logic operation stage (where $C$ is the capacitance between
the shorted pair of electrodes and the n$^+$-Si substrate). We will address the indirect dissipation
later, but first we address the direct contribution.

The energy dissipated due to Gilbert damping in a magnet is given by \cite{behin-aein,sun2005}
\begin{equation}
E_{d\textunderscore gd} = \int\limits_0^{t_s} P_{d\textunderscore gd} (t) dt,
\end{equation}
where $t_s$ is the switching delay (counted between the time the magnetization leaves the
vicinity of $\Psi_1$ and arrives within 4$^{\circ}$ polar angle of $\Psi_0$)
and $P_{d\textunderscore gd} (t)$ is the power dissipation and can be expressed as
\begin{equation}
P_{d\textunderscore gd} (t) = \frac{\alpha \gamma}{(1+ {\alpha}^2) \mu_0 M_s \Omega}
{\lvert \mathbf{{\tau}_{eff}}(t) \rvert}^2 ,
\end{equation}
where $\mathbf{\tau_{eff}}(t)$ is the vector sum  of the torques due to shape anisotropy, stress anisotropy and bias
 magnetic field (the thermal torque does not dissipate energy). The energy dissipation
is obviously different for different switching trajectories because of the integration over time, and we have found that the mean dissipation
is 316 kT at room temperature. This calculation overestimates the energy dissipation slightly, but that
only makes our figures conservative.

The next component is the $C \left ( V_{MN} \right )^2$ dissipation. We have electrodes of dimensions
80 nm $\times$ 80 nm and the thickness of the PZT layer is 50 nm. Thus, the
capacitance between either electrode and the silicon substrate
is C = 1.13 fF, assuming that the relative dielectric constant of PZT is 1000.
The maximum value of the voltage $V_{MN}$ is $V_0/3$ (= 56.3 mV). Since we have a {\it pair} of electrodes,  the
dissipation will be roughly twice (1/2)$C \left ( V_{MN} \right )^2$. We calculated that value as 865 kT.
Therefore, the total direct disssipation is at most 316 + 865 = 1181 kT = 4.9 aJ

\section{Charging time}

The output of one stage charges the input of the next stage. We need to calculate the time it takes
for the input of the next stage to charge up after the output of the preceding state changes. This should be
much shorter than the gate delay.

The PZT thin film of the successor gate acts as a capacitor whose one end is connected to bias voltage source $V_{BIAS}$
and the other is connected via
passive resistors $R$ and $R_P$ to the output of the preceding stage.
The equivalent charging circuit is shown in the left panel of Fig. 7
where C$_1$ represents the capacitance due to the PZT layer underneath the MTJ in the preceding
gate and C$_2$ is the capacitance due to the
PZT layer underneath the metal electrodes in the successor gate. The resistor R$_{MTJ}$ represents the resistance of
the MTJ stack of the preceding stage. The other resistors are the passive resistances at the input of the successor gate.

\begin{figure}
\centering
\includegraphics[width=6in]{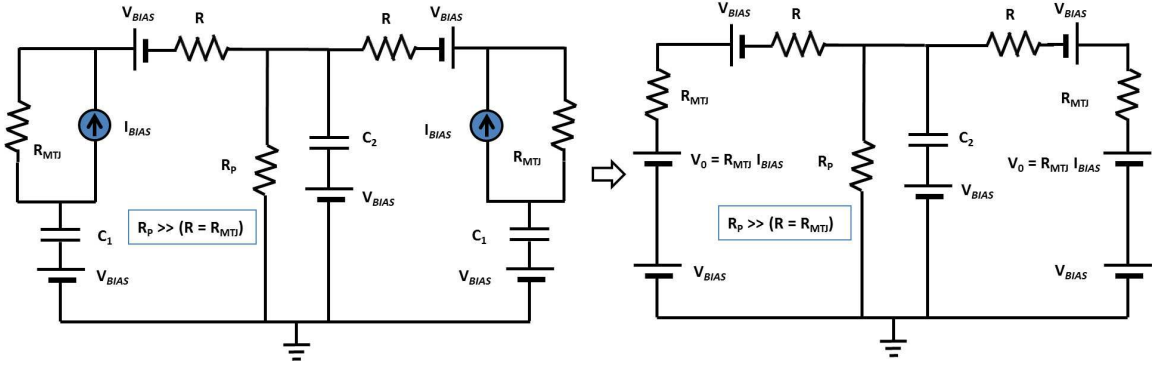} \label{charging}
\caption{Norton's equivalent circuit on the left and Thevenin's equivalent circuit on the right}
\end{figure}

In the right panel of the Fig. 7, we have converted the
Norton's equivalent circuit for the current source to Thevenin's equivalent circuit for a voltage source.
We have shorted the capacitor C$_1$ for the following reason. There is no electric field underneath the MTJ and
the voltage drop across
this region is almost zero. Since the initial and final voltages on this capacitor are zero, the charging/discharging of
this capacitor can be ignored and we will treat it as a virtual short.

The charging time constant of the circuit will therefore be
\begin{equation}
\tau = R_{eq} C_2; ~~~R_{eq} = R_P \parallel \left ( R + R_{MTJ} \right ) \parallel \left ( R + R_{MTJ} \right ) .
\end{equation}
We choose $R_P \gg R + R_{MTJ}$, which makes R$_{eq}$ $\approx$ (R + R$_{MTJ})/2$. We then select $R = R_{high} = R_0$
= 80 k$\Omega$
and R$_P$ = 10 G$\Omega$. This makes the charging time constant $\tau$ $\leq$ 180 picoseconds
where C$_2$ is 2$\times$1.13 fF (due to two metal electrodes). Compare this time with the magnet switching time of
1.3 ns. Since this time is much shorter than the magnet switching time, we can approximate the stress application
and withdrawal as virtually `instantaneous'.

\section{The bias current}
The bias current is given by $I_{BIAS} = V_0/R_{MTJ}$ = 0.1689/80 k$\Omega$ = 2.11 $\mu$A. This
current is passed through the soft magnet whose lateral area is $(\pi/4) \times 100 \times 42$ nm$^2$,
resulting in a current density of 0.064 MA/cm$^2$, which is far below the critical current for switching
a magnet due to spin transfer torque or domain wall motion. Therefore, the bias current source cannot cause
any unwanted spurious switching.

\section{Indirect energy dissipation}
Fig. 8, derived from Fig. 7, shows the equivalent circuit for calculating the indirect power dissipation.
 Once charged up, the capacitor $C_2$ becomes an open circuit and the equivalent voltage source V$_0$
(due to the bias current source $I_{BIAS}$) now drives
a current through the resistors, causing power dissipation that does not aid the switching, but is unfortunately 
unavoidable.

\begin{figure}
\centering
\includegraphics[width=6.5in]{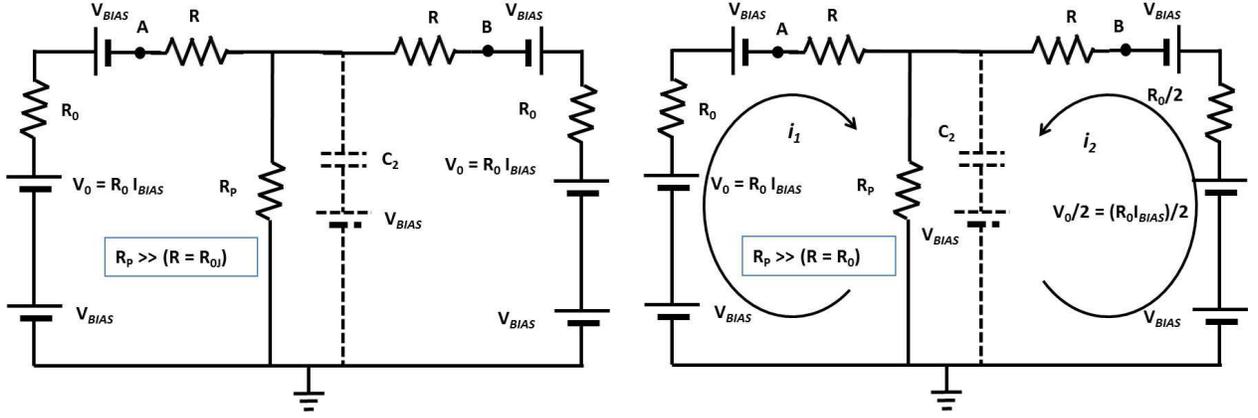} \label{stbpower}
\caption{Equivalent circuits for standby power dissipation when both inputs at node A and node B are high (left) and when either input is low (right).}
\end{figure}
If both inputs are at the same state (i.e. both are high, both are low, or in RESET) the voltage sources 
(both $V_0$, $V_0$/2, or $V_0$/4) drive  a current $i_1 = i_2 = I$ through the series resistors $R$ and $R_{MTJ}$ and a current 2$I$
through the parallel resistance $R_P$ where $I = V/\left (R + R_{MTJ} + 2R_P \right )$. This causes power dissipation
given by
\begin{equation}
P_{indirect1} = 2I^2 (R + R_{MTJ}) + 4I^2 R_P \approx \frac{V_0^2}{R_P}~~~~ {\tt since}~~  R_P \gg R+R_{MTJ}.
\end{equation}
The maximum value of V$_0$ is 0.1689 V; hence,
 the indirect energy dissipation over the gate delay
is E$_{indirect1}$ = P$_{indirect1}$ $\times$ t$_S$ $\approx$ 3.7 zJ which is negligible (t$_S$ is the switching time).

When either input is low, one voltage source is at $V_0$ and other is at $V_0$/2. The MTJ resistances in the two branches 
are $R_{MTJ1} = R_{high} = R_0$ and $R_{MTJ2} = R_{low} = R_0/2$. Currents flow through the series resistors
$R$ and $R_0$ in the left loop, and $R$ and $R_0/2$ in the right loop, as well as the parallel resistor $R_P$, as shown in the right panel of
Fig. 8. By solving the circuit in the right panel, we
find that the currents {\it i$_1$} and {\it i$_2$} are 301.612 nA and -301.6 nA respectively. The corresponding power
dissipations in
$R + R_{0}$ due to {\it i$_1$}, and in $R + R_{0}/2$ due to {\it i$_2$}, are 14.55 nW and 10.9 nW, respectively.
The current that
flows through the resistor R$_P$ is $i_1 + i_2$, which is very small (12 pA) and hence the power dissipation in this
resistor is negligible. Therefore, the total indirect energy dissipation
E$_{indirect2}$ is (14.55 + 10.9) nW $\times$ 1.3 ns = 33.08 aJ (7986 kT).
We can clearly see that E$_{indirect2}$ is much larger than E$_{indirect1}$ when either input is low. Fortunately,
in a random input stream, this situation arises 50\% of the time. Hence, the time averaged energy dissipation will be
16.54 aJ or 3993 kT.

Consequently, in one cycle of a gate operation, the time-averaged energy dissipation is roughly 4.9 + 16.54 = 21.44 aJ = 5176 kT.
The time averaged energy-delay product is 21.44 aJ $\times$ 1.3 ns = 2.78$\times$10$^{-26}$ J-sec.
 This figure is almost two orders of magnitude smaller than that of any other magnetic non-volatile logic gate proposed
 until now (to our knowledge) and one order of magnitude smaller than that of CMOS gates.

\section{A few fundamental considerations}

This logic gate is switched by charging a capacitor $C_2$ which generates strain in a piezoelectric
and that strain switches the MTJ resistance. Charging a capacitor involves moving an amount of charge $\Delta Q$
  in and out of a region in some time $\Delta t$, resulting in a charging current $I = \Delta Q/ \Delta t$ and an
 associated energy dissipation
$E_d = I^2 R \Delta t = (\Delta Q)^2 R/\Delta t = \Delta Q \Delta V$ where $\Delta V = I R$ \cite{bandy_nanoscience}.
The energy-delay product is $E_d \times \Delta t = (\Delta Q)^2 R$. In our gate, the quantity $\Delta Q = C_2 V_{MN}$
which is equal to
is 2 $\times$ 1.13 fF $\times$  56.3 mV = 0.127 fC = 795 electronic charge. In contrast, when magnets are switched with spin polarized
current directly, the amount of charge moved is at least $\sim$20,000 \cite{Datta-review}. This may portend a fundamental
advantage of voltage-controlled straintronic switching over current-induced switching of magnets.

The energy-delay product $(\Delta Q)^2 R$ should be  (0.127 fC)$^2$ $\times$ 80 k$\Omega$ = 1.3$\times$10$^{-27}$ J-sec. This is
the energy-delay product associated with just charging and discharging the gate in response to inputs. There are additional
sources of dissipation in the gate due to internal losses and additional delay due to magnet switching. These additional
components make the actual energy-delay product roughly an order of magnitude larger in our case.

\bigskip

{\small

\noindent {\bf Acknowledgement}

This work is supported by the US National Science Foundation (NSF) under grants
ECCS-1124714 and CCF-1216614 and by Semiconductor Research Company (SRC) under NRI task 2203.001. J. A's work is also supported by NSF CAREER grant CCF-1253370.

\bigskip

\noindent {\bf Author contributions}

A. K. B. carried out the stochastic Landau-Lifshitz-Gilbert
simulations to evaluate the error probability. S. B and J. A. came up
with the idea
and verified the simulation results. All authors contributed to writing the paper.

\bigskip

\noindent {\bf Competing financial interests}

The authors declare no competing financial interests.}

\end{document}